\documentstyle[11pt,epsfig]{article}
\title{
 Transverse Spin Structure Function $g_{2}(x,Q^2)$ in the Valon model
 }

\author{ Z.Alizadeh Yazdi$^{(1)}$\footnote{zahra.alizadehyazdi@stu.um.ac.ir}, A.Tahamtan $^{(2)}$\footnote{azamtahamtan2@yahoo.com},
 F.Taghavi-Shahri$^{(1)}$\footnote{taghavishahri@um.ac.ir}, F.Arash $^{(3)}$\footnote{farash@cic.aut.ac.ir} \\
 and M.E.Zomorrodian $^{(1)}$\footnote{ zomorrod@ferdowsi.um.ac.ir}
\\
$^{(1)}$ Department of Physics, Ferdowsi University of Mashhad,
P.O. Box 1436,\\ Mashhad, Iran\\
$^{(2)}$ Department of Physics, University of Tehran, North Kargar Avenue,\\
 Tehran 14395-547, Iran\\
$^{(3)}$ Physics Department, Tafresh University, Tafresh, Iran\\
 }
\date{\today}
\begin{document}
\maketitle
\begin{abstract}
The spin dependent structure function, $g_2^{ww}$, is calculated
in the valon model. A simple approach is given for the
determination of the twist-3 part of the $\bar{g_{2}} (x,Q^2)$ in
Mellin space; thus, enabling us to obtain the full transverse
structure function, $g_2 (x,Q^2)$ for proton, neutron and the
deuteron. In light of the new data, we have further calculated the
transversely polarized structure function of
$g_{2}^{3_{He}}(x,Q^2)$. Our results are checked against the
experimental data and nice agreements are observed.

\end{abstract}


\section{INTRODUCTION}

The nucleon polarized structure functions $g_{1,2}(x,Q^2)$ are
important tools in understanding the nucleon substructure. In
particular, they are indispensable elements for the understanding
of the spin dependent parton distributions and their correlations.
The $g_{2}(x,Q^2)$ structure function is important because it
probes transversely and also longitudinally polarized parton
distributions inside the nucleon. $g_{2}(x,Q^2)$ structure
function is also sensitive to the higher twist effects, such as
quark gluon correlations. They do not disappear even at large
$Q^2$ values and  not easily interpreted in pQCD
\cite{shuryak1,shuryak2}. Since $g_{2}(x,Q^2)$ is the only
function related to the quark-gluon interaction, learning
about its behavior will render further insight into the spin structure of the nucleon beyond the simple quark parton model.\\
Thus, the main purpose of this paper is to calculate transverse spin structure function,
$g_{2}(x,Q^2)$. As such, it requires considering both the twist-2 and the twist-3 contributions.
Here we will present a simple method to extract twist-3 part. The twist-2 part is
 well understood, and requires knowledge about the $g_{1}(x,Q^2)$ structure function.
 Therefore, first we will briefly  review $g_{1}(x,Q^2)$ in the context of the so called valon model representation of hadrons.\\
Finally, the outcome of our results is checked against the
experimental data from
\cite{abe,Anthony,E155x,HERMES,Airapetian}, and compared with other phenomenological models.\\
The lay out of the paper is as follows: In section 2, we briefly
present a review of  the polarized nucleon structure function in
the valon model. Section 3 deals with the calculation of
$g_{2}(x,Q^2)$ spin structure function and discusses the numerical
results. We also provide some discussion on the effect of higher twists. Section 4 is devoted to the sum rules. Our conclusions are given in section 5.\\

\section{A brief review of spin structure functions in the valon model }

The valon model is a phenomenological model, originally proposed
by R. C. Hwa, \cite{Hwa1} in early 80's to provide a bridge
between the naive quark model and the partonic structure of the
hadrons. The model had many successes. It was improved later by
Hwa \cite{Hwa2} and Others
\cite{Altarelli,Arash1,Arash2,Arash3,Arash4,Arash5}. It was
further extended to include the polarized cases \cite
{Arash6,Arash7,Arash8,Arash9}. The model views a hadron as three
(two) constituent quark like objects called valons. Each valon is
defined to be a dressed valence quark with its own cloud of sea
quarks and gluons. The dressing processes are
described by QCD. At high enough $Q^2$ values the structure of a valon can be resolved, but At low $Q^2$ values, the internal structure of the valon cannot be resolved and it behave as a constituent quarks of the hadron.\\
In valon model the polarized parton distributions of a polarized
hadron are given by the following convolution integral :
\begin{equation}\label{deltaq}
\delta q_{{i}}^{\it{h}}(x,Q^2)=\sum \int_{x}^{1}
\frac{dy}{y}\delta G_{\it{valon}}^{h}(y)  \delta
q_{{i}}^{\it{valon}}(\frac{x}{y},Q^2)
\end{equation}
where $\delta G_{\it{valon}}^{h}(y)$ is the valon helicity distribution
in the hosting hadron; that is, it is the probability of finding a polarized
valon inside the polarized hadron. In the Next-to-Leading order, $\delta G_{\it{valon}}^{h}(y)$ marginally depends on $Q^2$. These distributions are shown in (Fig.1).

\begin{figure}[htp]\label{udvalon}
\centerline{\begin{tabular}{cc}
\includegraphics[width=8.5cm]{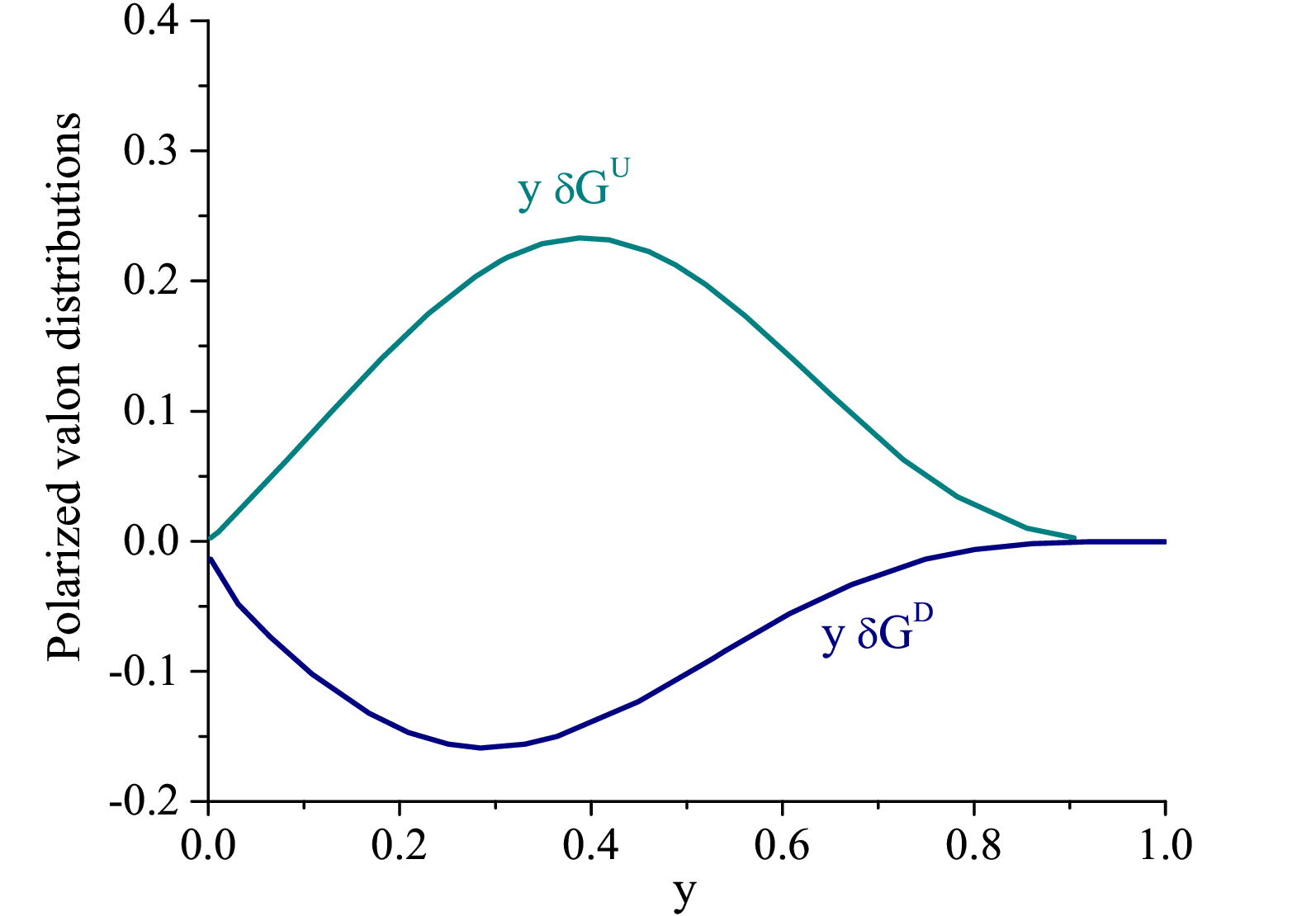}
\end{tabular}}
 \caption{\footnotesize
(Color online)  polarized valon distribution functions
 for U and D valon types (the
helicity distributions for the valons) inside the proton.   }
\end{figure}

The term $ \delta q_{{i}}^{\it{valon}}(x/y,Q^2)$ in
Eq.(\ref{deltaq}) is the polarized parton distribution (PPDFs)
inside a valon and are obtained from the solutions of DGLAP
evolution equations in the valon. Now, using the convolution
integral, one can obtain the polarized hadron structure functions
as follows:
\begin{equation}\label{gone}
g^{h}_{1}(x,Q^{2})=\sum_{\it{valon}}\int_{x}^{1}\frac{dy}{y}
\delta G_{\it{valon}}^{h}(y) g^{\it{valon}}_{1}(\frac{x}{y},Q^{2})
\end{equation}
where $g^{\it{valon}}_{1}(\frac{x}{y}, Q^{2})$ is the polarized
structure function of the valon. The details of actual
calculations are given in \cite{Arash6,Arash8}. In short, the
following two steps lead us to both the polarized PPDFs and
polarized nucleon structure functions:
\begin{itemize}
\item Calculate the PPDFs in the valon using DGLAP equations;
\item With a phenomenological approach, the helicity distributions
of the valons in a nucleon is obtained. then, they are used in
Eqs.(\ref{deltaq}) and (\ref{gone})to get the polarized parton
distribution functions (PPDFs) and the polarized nucleon
structure up to $Q^2=10^7 GeV^2$;

\end{itemize}
It should be noted that the valon model is only a phenomenological
model. As such, initial conditions, as inputs to the DGLAP
equations, are chosen based on phenomenological arguments. The
results obtained for the proton structure function, $g_{1}^{p} (x,
Q^2)$ from this model are in excellent agreement with all
available experimental data
\cite{COMPASS1,Airapetian2,COMPASS2,Kramer,abe2,abe3}. In Fig.2 we
only present a sample of the results along with the existing data.
The parton distributions so obtained will be used here to
calculate $g_{2} (x, Q^2)$.

\begin{figure}[htp]
\centerline{\begin{tabular}{cc}
\includegraphics[width=7.5cm]{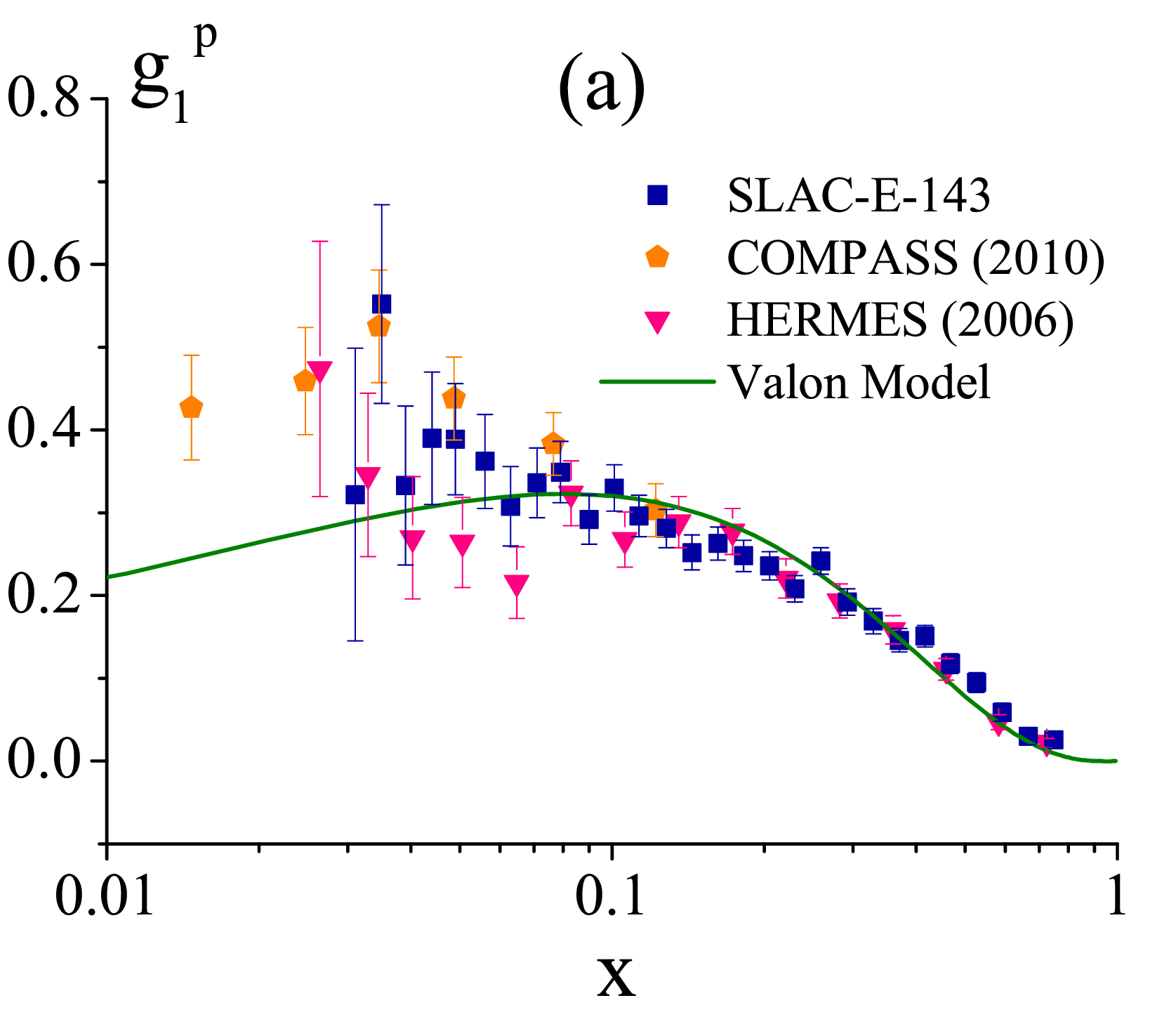}
\end{tabular}}
\centerline{\begin{tabular}{cc}
\includegraphics[width=7.5cm]{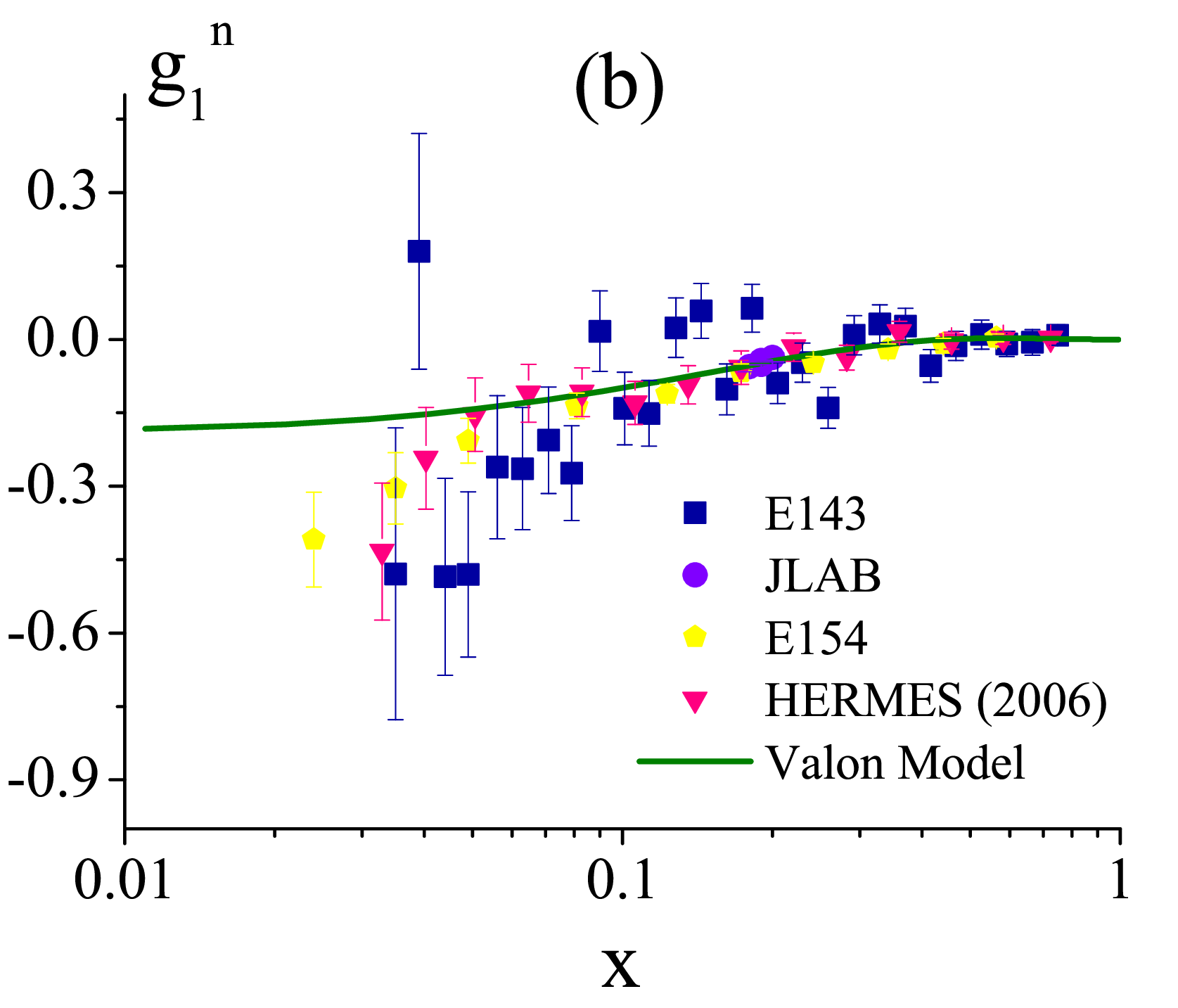}
\end{tabular}}
\centerline{\begin{tabular}{cc}
\includegraphics[width=7.5cm]{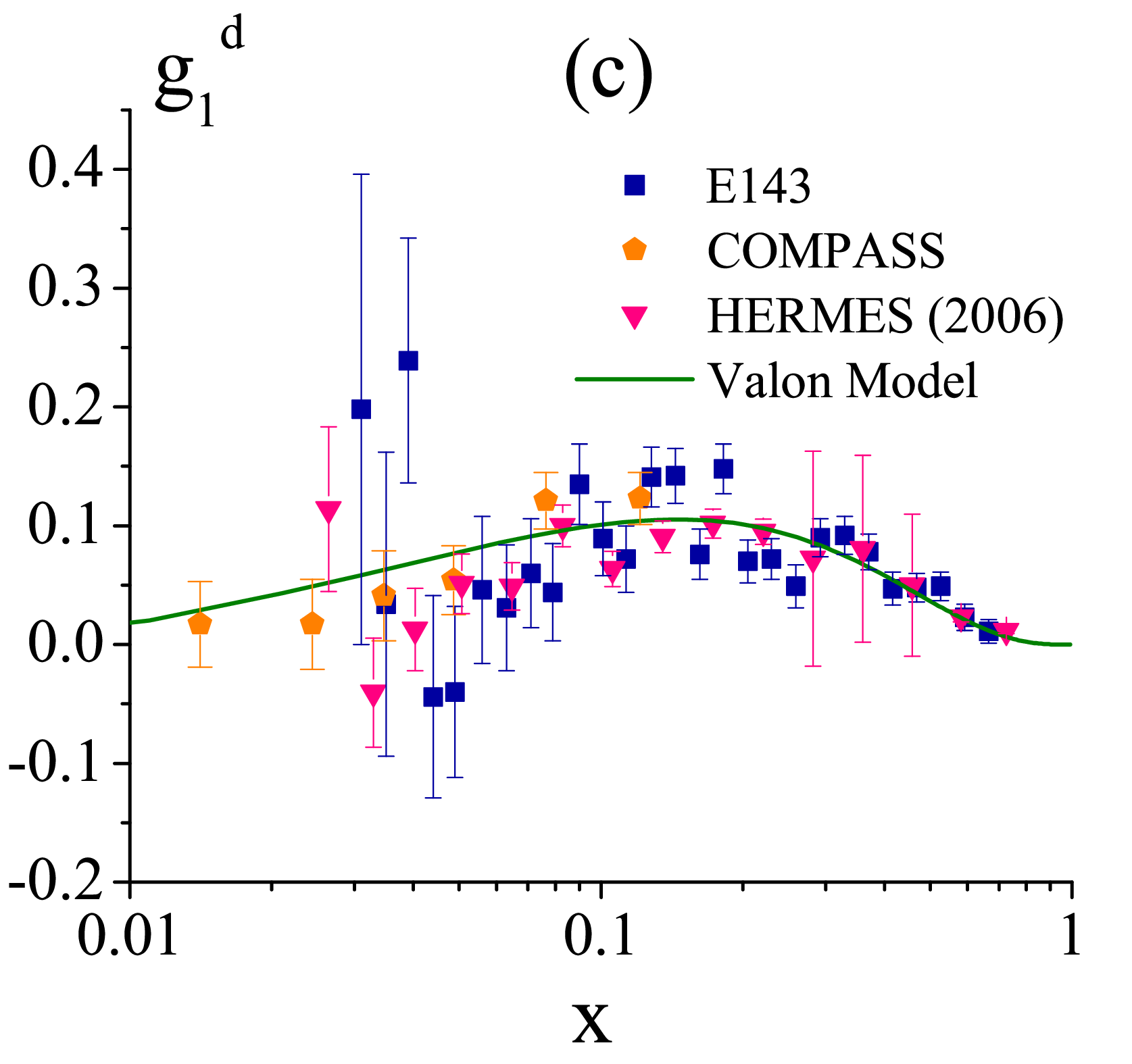}
\end{tabular}}
 \caption{\footnotesize   (Color online) Polarized proton, neutron and deuteron
structure function $g_{1}(x,Q^2)$ at $Q^2=5\ GeV^2$ . The results
from valon model  are compared with the experimental data
\cite{COMPASS1,Airapetian2,COMPASS2,Kramer,abe2,abe3}. }
\end{figure}

\section{Transverse spin-dependent structure function $g_{2}(x,Q^2)$ }

Polarized Deep Inelastic Scattering (DIS) mediated by a photon
exchange, probes two spin structure functions: $g_{1}(x,Q^2)$ and
$g_{2}(x,Q^2)$. If the target is transversely polarized, the total
cross section is a combination of these two structure functions.
Transverse spin structure function, $g_{2}(x,Q^2)$, is made up of
two components: a twist-2 part, $g_2^{ww}$, and a mixed twist
part, $\bar g_{2}(x,Q^2)$. Therefore, it can be written as
\cite{Cortes}:
\begin{equation}\label{gtwo}
g_{2}(x,Q^2)=g_{2}^{ww}(x,Q^2)+\bar g_{2}(x,Q^2)
\end{equation}
 where
\begin{equation}\label{gbar}
\bar g_{2}(x,Q^2)=-\int_{x}^{1}\frac {\partial}{\partial
y}(\frac{m}{M} h_T(y,Q^2)+\xi(y,Q^2))\frac{dy}{y}.
\end{equation}
The twist-2 part, $g_2^{ww}$, comes from OPE. The $\bar g_{2}
(x,Q^2)$ receives a contribution from the transversely polarized
quark distributions $h_T(x,Q^2)$ plus a contribution that
comes from a twist-3 component, an indication of $qgq$ correlations, given by
$\xi(y,Q^2)$ term in Eq.(\ref{gbar}). These higher
twist corrections arise from the non-perturbative multi parton
interactions. Their contributions at low energy increase as $\frac
{1}{Q^\tau}$, reflecting the confinement. Any non-zero result for
this term at a given $Q^2$ will reflect a departure from the
non-interacting partonic regime \cite{RSS}.\\
$g_2^{ww}$ is related to the $g_{1}$ structure function by the
Wandzura-Wilczek relation \cite{Wandzura} as follows,
\begin{equation}\label{gww}
g_{2}^{ww}(x,Q^2)=-g_{1}(x,Q^2)+\int_{x}^{1}g_{1}(y,Q^2)\frac{dy}{y}.
\end{equation}

\subsection{Calculation of the twist-2 term, $g_2^{ww }(x,Q^2)$}

We begin with Eq.(\ref{gww}). Since $g_{1}(x,Q^2)$ is known in the
valon model \cite{Arash6,Arash8}, we utilize them without any
additional free parameter and evaluate the twist-2 part of
$g_2(x,Q^2)$; namely $g_2^{ww }(x,Q^2)$, according to the
Eq.(\ref{gww}). The results are shown in Fig. 3 for proton. We
have also included the findings of \cite{Song1,Song2} for the
purpose of comparison.
\begin{figure}[htp]
\centerline{\begin{tabular}{cc}
\includegraphics[width=7.5cm]{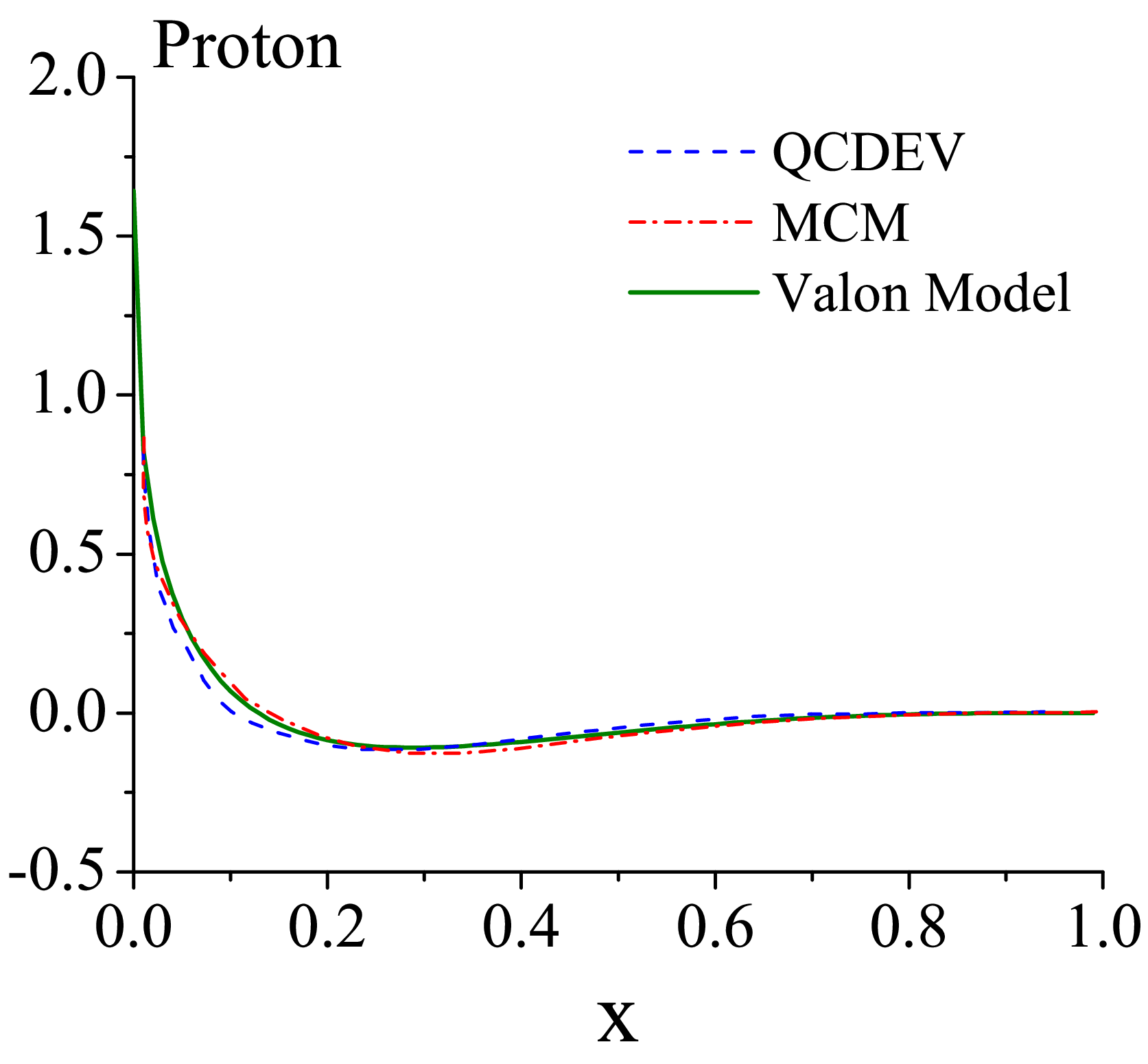}
\end{tabular}}
 \caption{\footnotesize  (Color online) Transverse polarized proton
structure functions, $g_2^{ww}(x,Q^2)$ at $Q^2=5\ GeV^2$. We
compare our results with the results from other phenomenological
models \cite{Song1,Song2}.}

\end{figure}

\subsection{Calculating the twist-3 term, $\bar g_2(x,Q^2)$}
As mentioned, the function $\bar g_{2}(x,Q^2)$ has two
terms. The first term is a twist-2 contribution related to the
transverse polarization of quarks in the nucleon. It is suppressed
by the quark to nucleon mass ratio and will be ignored here due to
its negligibility. The second part is a twist-3 contribution, reflecting
the quark-gluon correlations. In the following we will focus on this part. \\
In large $N_{c}$ limit, Ali, Braun and Hiller found that the
$Q^2$-evolution of $\bar g_{2}(x,Q^2)$ is qualified by simple
DGLAP type equation with awhile difference between the anomalous
dimensions and the twist-2 distribution\cite{Ali,Wakamatsu}. It implies that
$\bar g_2(x,Q^2)$ obeys the following simple equation :
\begin{equation}\label{gmellin}
\bar g_2(n,Q^2)=L^{\frac {\gamma _n^g}{2 b_0}}\bar g_2(n,Q_0^2)
\end{equation}
where,
\begin{equation}
\bar g_2(n,Q^2)=\int_{0}^{1} x^{n-1} \bar g_2(x,Q^2)dx,
\end{equation}
\begin{equation}\label{l}
L\equiv \frac {\alpha _s(Q^2)}{\alpha _s(Q_0^2)},\\
\end{equation}
and
\begin{eqnarray}
b_0=\frac {11}{3} N_c - \frac {2}{3}N_f\\
\gamma _n^g=2 N_c (S_{n-1}-\frac {1}{4}+\frac {1}{2 n})\\
S_{n-1}=\sum \frac {1}{j}
\end{eqnarray}
$\alpha _s$ is the strong coupling constant, $N_{f}$ is the number
of flavor and $S_{n-1}$ are the Harmonic functions. Our purpose is
to find the $Q^2$-evolution of $\bar g_2$ with some appropriate
initial conditions in moment space. Then we can make a
transformation to the momentum space and evaluate the twist-3
contribution to the transverse spin structure function. This is
done in two steps, as is the case in the valon model. The first
step involves finding a solution to Eq.(\ref{gmellin}) in a valon.
The second step is to convolute the results obtained in the first
step with the valon distribution in the nucleon. This will give
the nucleon structure function.\\
We take $Q_{0}^{2} = 0.238 $ as our initial scale which is also
used in our original calculations of various parton distributions.
This value of $Q_{0}^{2}$ corresponds to a distance scale of $0.36
fm$ which is roughly equal to or less than the radius of a
valon\cite{Arash1}. The initial input function for $\bar
g_2^{valon}(z,Q_0^2)$ is (See the Appendix A) :
\begin{equation}\label{g22valon}
 \bar g_2^{valon}(z,Q_0^2) = A  \delta (z-1)
\end{equation}
The justification for this choice is as following: In the momentum
space one can write
\begin{equation}
\bar g_2^{valon}(z,Q^2) =f(Q^2)\bar g_2(z,Q_0^2)=f(Q^2) A \delta
(z-1),
\end{equation}
Note that for $Q^2 =Q_0^2$ we get $f(Q^2)\rightarrow f(Q_0^2)=1$
which is apparent from the definition of $L$ in Eq.(\ref{l}),
thus, arriving at Eq.(\ref{g22valon}). This simple choice for the
initial input in $\bar g_2^{valon}(z,Q_0^2)$ stems from the
knowledge that it is related to the quark gluon correlations,
which in turn, is related to the Green function in the momentum
space. In the momentum space the correlation function is composed
of a Dirac Delta term and a function that is related to the
momentum. consequently, at the initial $Q_0^2$ we can simply
assume that $\bar g_2^{valon}(z,Q_0^2)$ is proportional to Dirac
Delta function which emphasizes the conservation of energy-
momentum and the fact that at such a low $Q_0^2$ a valon behaves
as an object without any internal structure. The last point is
built in the definition of a valon. So, in the moment space the
Delta function becomes unity and we can write
\begin{equation}\label{a}
\bar g_2(n,Q_0^2) = A \times 1
\end{equation}
all of the QCD effects are summarized in $A $. This coefficient
will be extracted from the experimental data. A fit to E143 data
yeilds a value $0.01$ for A. Having specified the initial input
values, the moments of $ \bar g_2^{valon}(z,Q^2)$  in a valon are
readily obtained, with the aid of Eq.(\ref{a}). An inverse Mellin
transformation then takes us to the momentum space, giving $\bar
g_2^{valon}(z,Q^2)$ structure function. For example at $Q^2=10
GeV^2$, we have:

\begin{equation}\label{g2valon}
 \bar g_2^{valon}(z,Q^2=10 GeV^2) = 0.028 z^{4.645}
\end{equation}
This completes the first step described above. In Fig.4,  $ \bar
g_2^{valon}(z,Q^2)$ is shown for different values of $Q^2$.
According to Fig.5 as our calculations for the values of
$N_{c}\geq 100$ leads to a similar distribution for $\bar
g_2^{valon}(z,Q^2)$ for the whole range of $z$, We are choosing
the optimal value of  $N_{c}$ which is equal to $100$ (reminder
that Eq. (6) is valid only for large $N_C$).

\begin{figure}[htp]
\centerline{\begin{tabular}{cc}
\includegraphics[width=7.5cm]{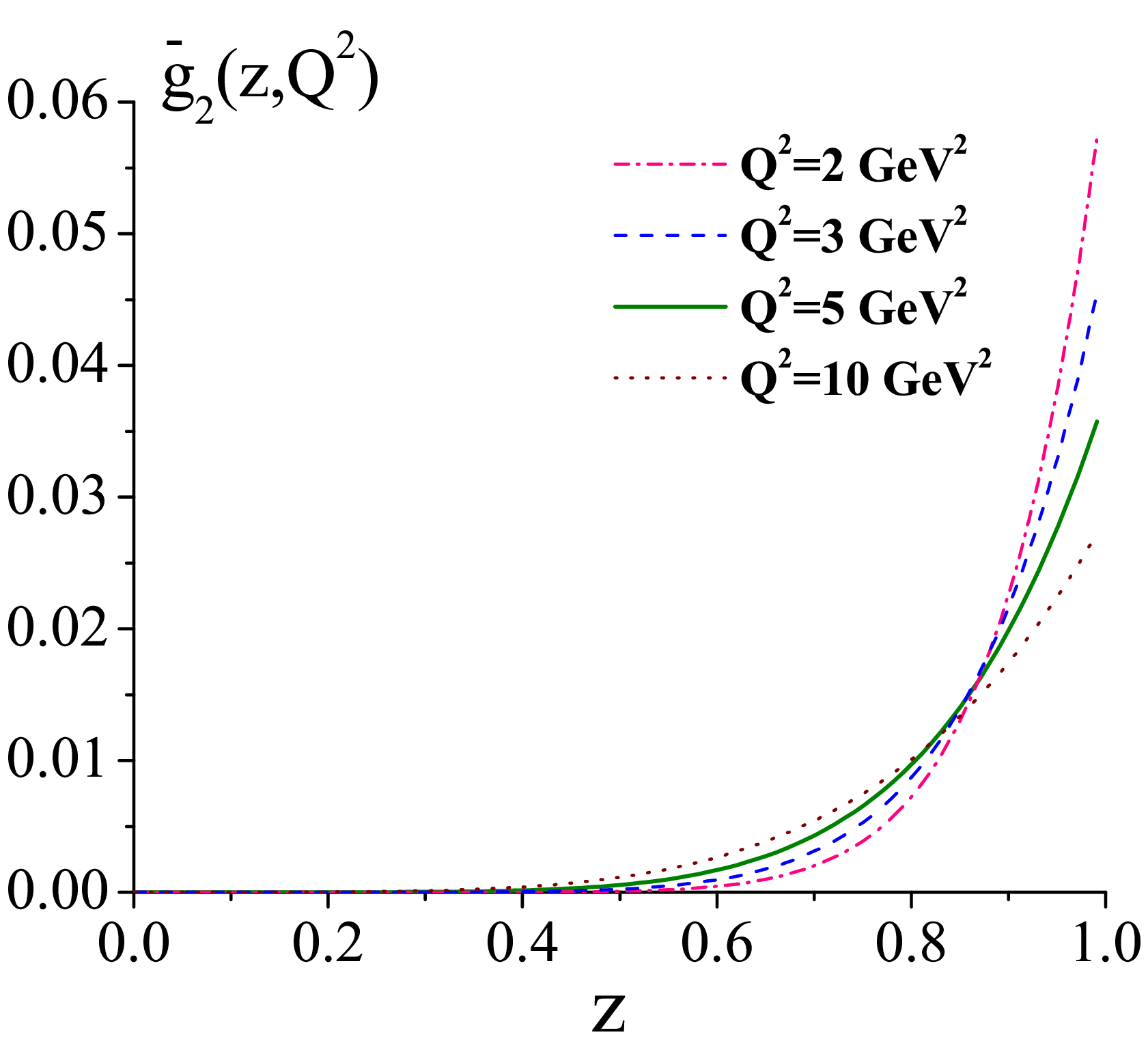}
\end{tabular}}
 \caption{\footnotesize (Color online) The twist-3 part of proton transverse spin structure function $ \bar g_2(z,Q^2)$ in the valon, at $Q^{2} = 2,3,5,10\ GeV^{2}$.}

\end{figure}

\begin{figure}[htp]
\centerline{\begin{tabular}{cc}
\includegraphics[width=7.5cm]{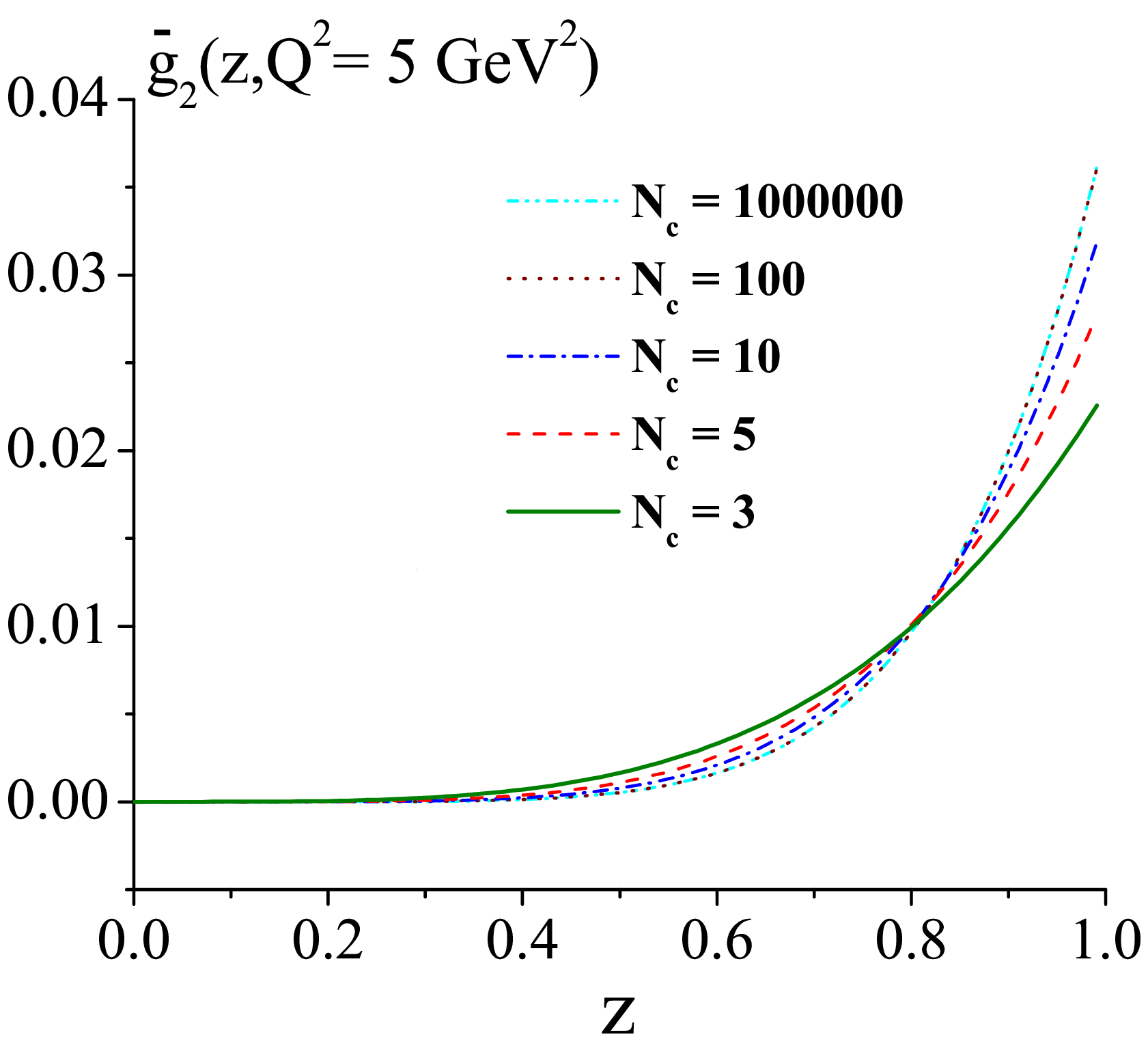}
\end{tabular}}
 \caption{\footnotesize (Color online) The twist-3 part of proton  transverse spin structure function $ \bar g_2(z,Q^2)$ in the valon  at $Q^{2} = 5\ GeV^{2}$ for different $N_{c}$.}

\end{figure}

The second step involves the convolution process which takes us to
the hadronic level. This is similar to the earlier procedure that
we have used to extract $g_1^{p,n}(x,Q^2)$. Similar to $g_{1}$, we
can write:
\begin{equation}\label{gtwoo}
g^{h}_{2}(x,Q^{2})=\sum_{\it{valon}}\int_{x}^{1}\frac{dy}{y}
\Delta_T G_{\it{valon}}^{p}(y)
g^{\it{valon}}_{2}(\frac{x}{y},Q^{2})
\end{equation}
where $ \Delta_T G _{valon}^{p} (y) $ is the transverse valon
distribution functions describing the probability of finding a
valon with spin aligned or anti-aligned with the transversly
polarized proton. In fact, $ \Delta_T G _{valon}^{p} (y) $ is
identical to $\delta G_{\it{valon}}^{p}(y) $ in the longitudinal
case. This is so,  because we know  that in the non-relativistic
limit of the quark motion, the PPDFs and transversity distribution
would be identical, since the rotations and Euclidean boosts
commute and a series of boosts and rotation can  convert a
longitudinal polarized proton into a transversely polarized one
with an infinite momentum \cite{alizadeh, Anselmino2009,ts1}.
Finally substituting the valon helicity distribution, $\delta
G^{U(D)}(y)$ in the corresponding hadron leads us to $\bar
g_2(x,Q^2)$. In Fig.6 we present $x^2 \bar g_2(x,Q^2)$ for both
proton and deuteron along with the E143 data \cite{abe} . We have
compared our results
with those obtained in the Bag model \cite{Song1} and that of Wakamatsu \cite{Wakamatsu}.\\
\begin{figure}[htp]
\centerline{\begin{tabular}{cc}
\includegraphics[width=7.0cm]{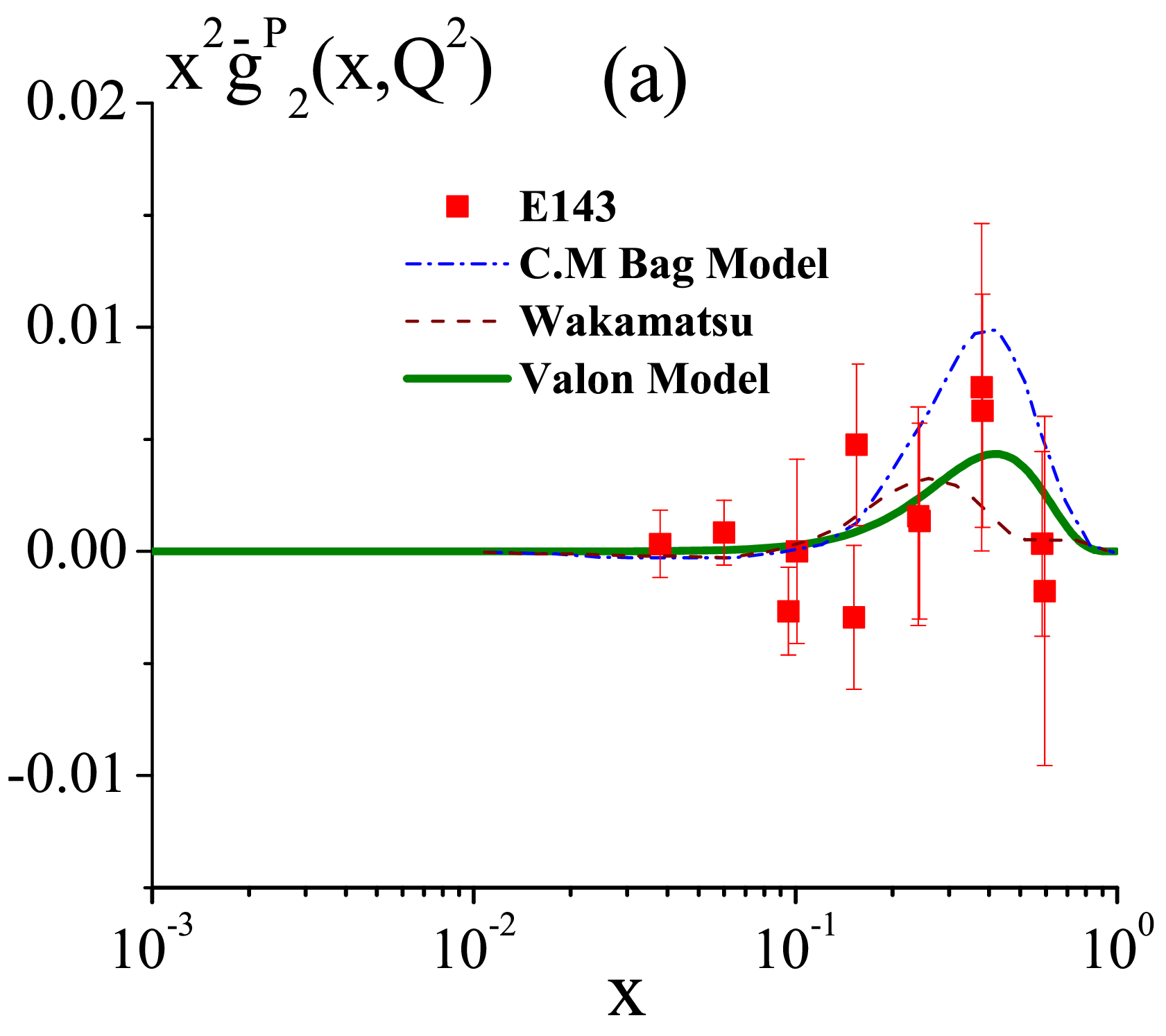}
\includegraphics[width=7.0cm]{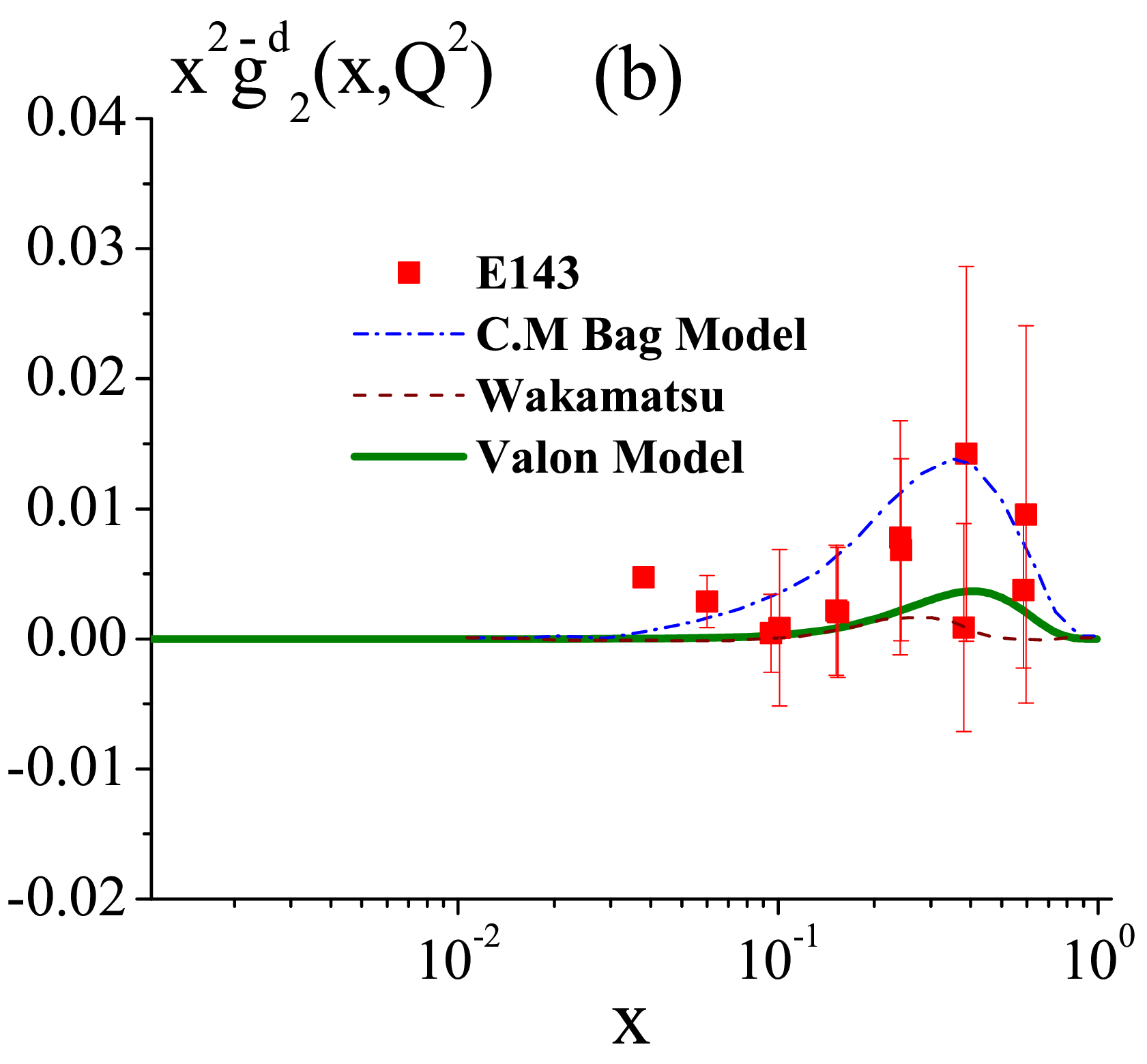}
\end{tabular}}
 \caption{\footnotesize  (Color online) The twist-3 part of proton and deuteron transverse spin structure function, $x^2 \bar g_2(x,Q^2)$ at
$Q^2=5\ GeV^2$. we compared our results with E143 data \cite{abe}
and other phenomenological groups  \cite{Song1}, \cite{Wakamatsu}.
 }
\end{figure}
Finally, adding $\bar g_2(x,Q^2)$ and $g_2^{ww}(x,Q^2)$, gives the
full $g_2(x,Q^2) $. The final results for $xg_{2}(x,Q^{2})$ are
presented in Fig.7 for proton, neutron and deuteron at different
values of $Q^{2}$.
\begin{figure}[htp]
\centerline{\begin{tabular}{cc}
\includegraphics[width=7.5cm]{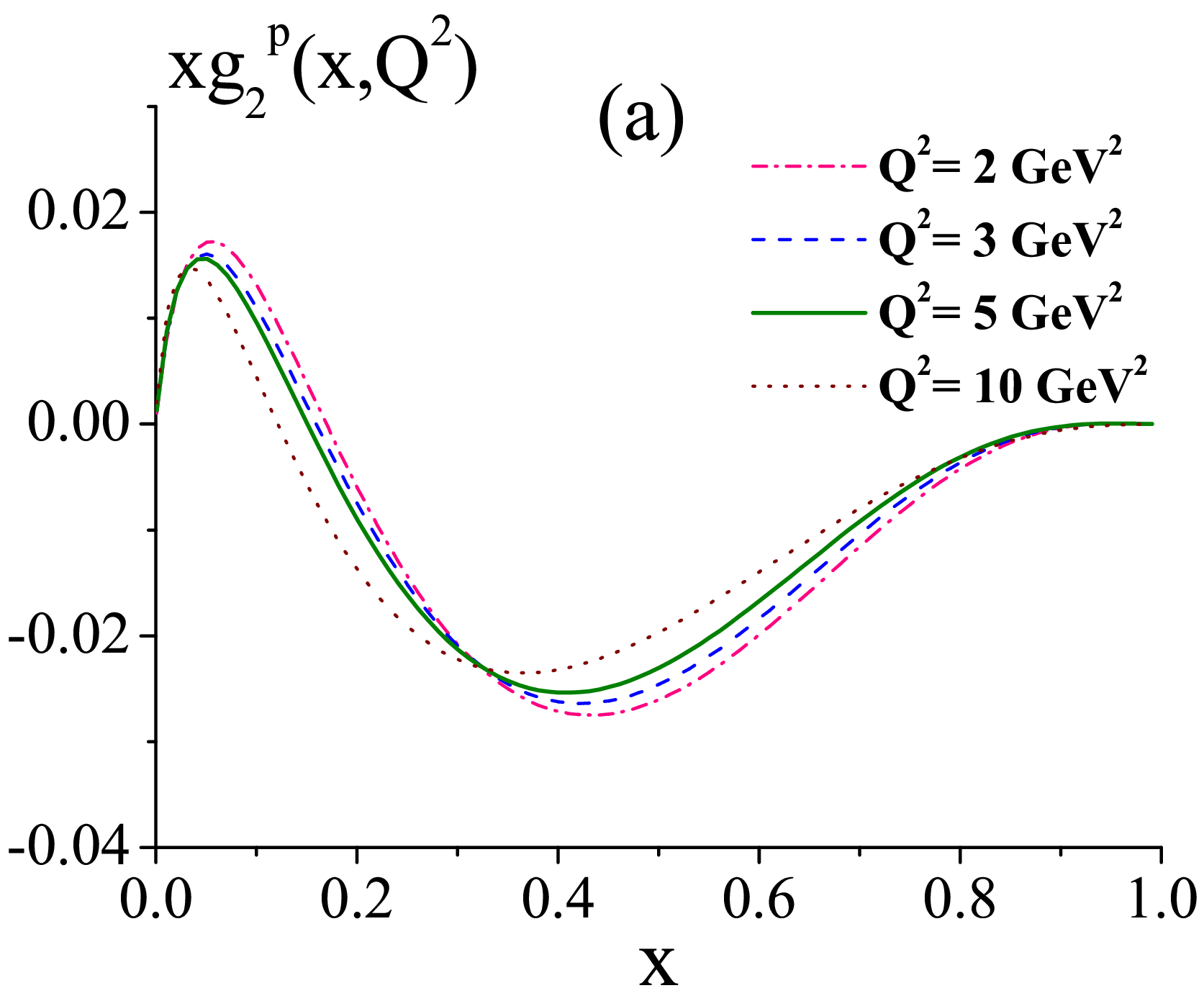}
\end{tabular}}
\centerline{\begin{tabular}{cc}
\includegraphics[width=7.5cm]{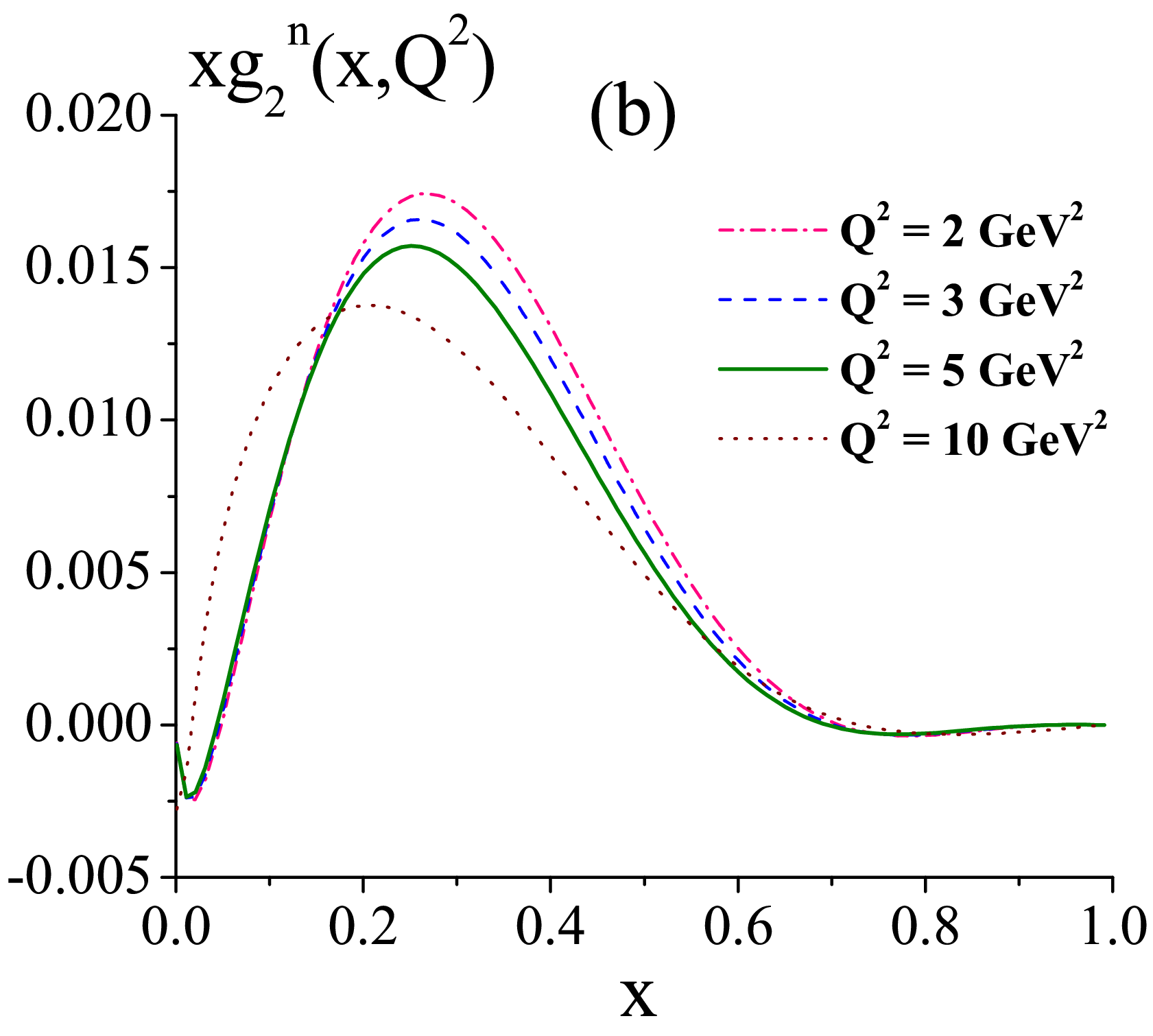}
\end{tabular}}
\centerline{\begin{tabular}{cc}
\includegraphics[width=7.5cm]{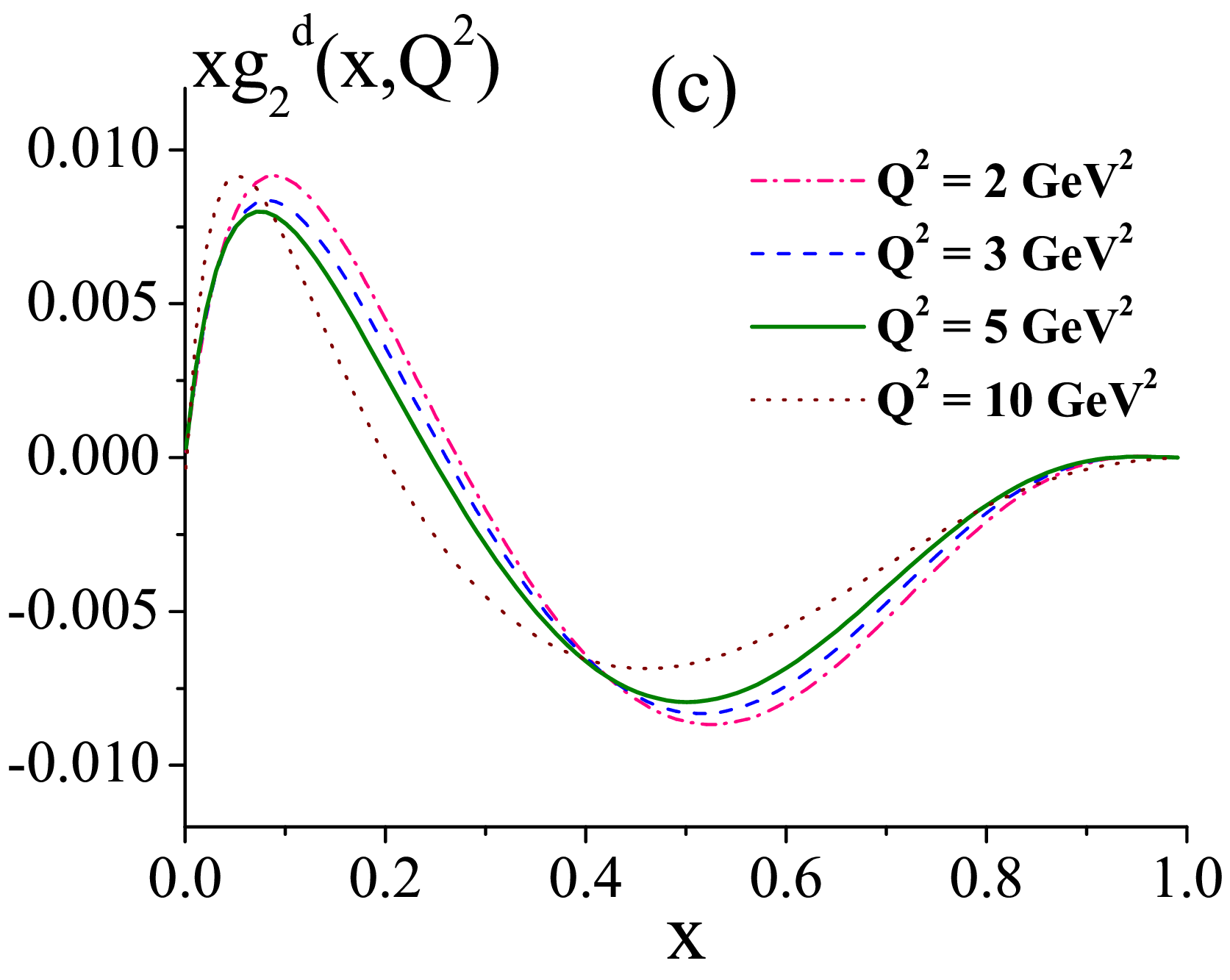}
\end{tabular}}
 \caption{\footnotesize  (Color online) $ xg_2(x,Q^2)$  proton, neutron and
 deuteron   at $Q^2 = 2,3,5,10\ GeV^2$
 .}
\end{figure}
Confronting with the experimental data, in Fig.8 we show our
results for the full transversely polarized structure function $
g_2(x,Q^2)$ for proton, neutron and deuteron and the experimental
findings of \cite{abe,Anthony,HERMES}.
\begin{figure}[htp]
\centerline{\begin{tabular}{cc}
\includegraphics[width=6.8cm]{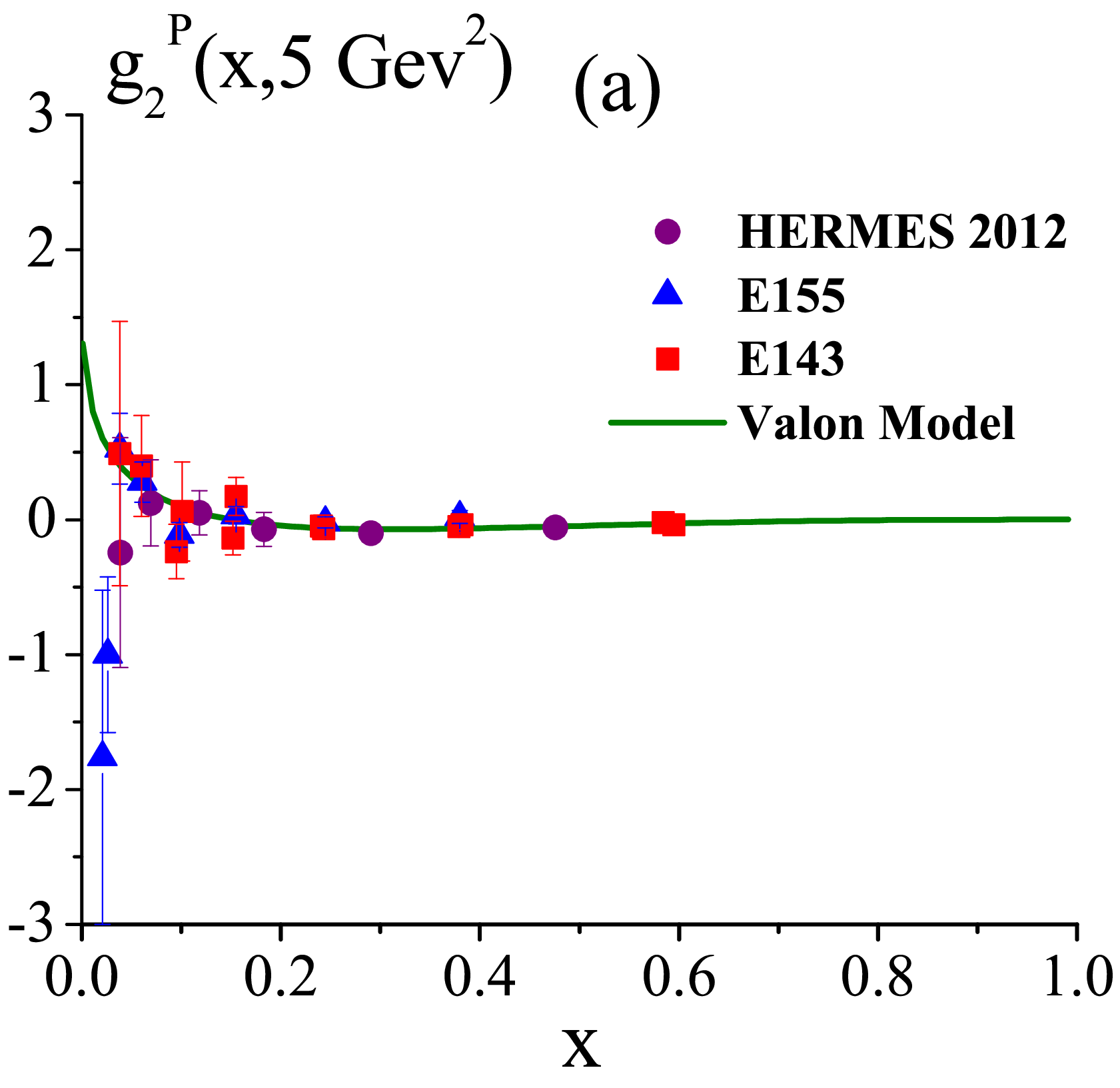}
\end{tabular}}
\centerline{\begin{tabular}{cc}
\includegraphics[width=6.8cm]{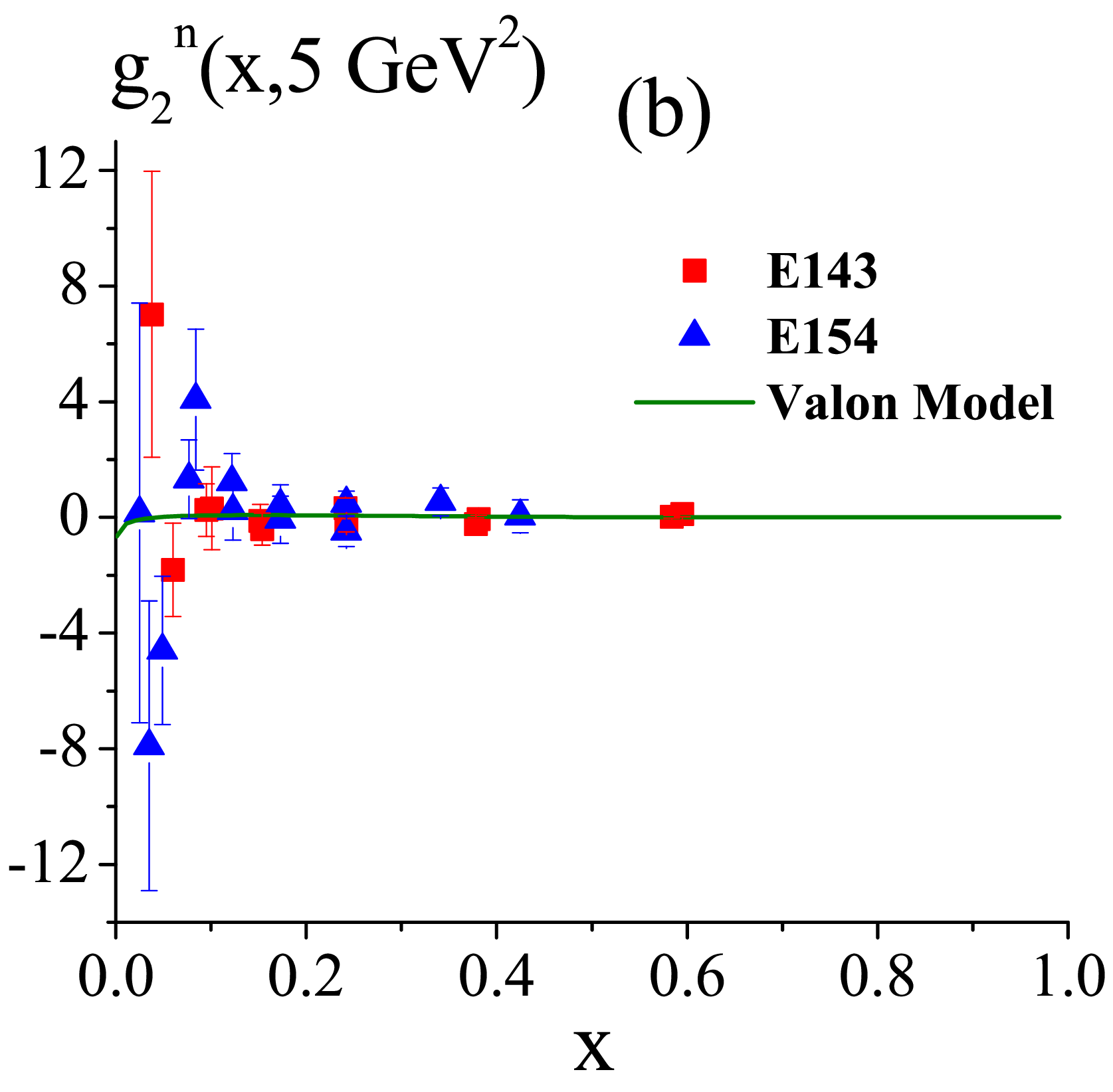}
\end{tabular}}
\centerline{\begin{tabular}{cc}
\includegraphics[width=6.8cm]{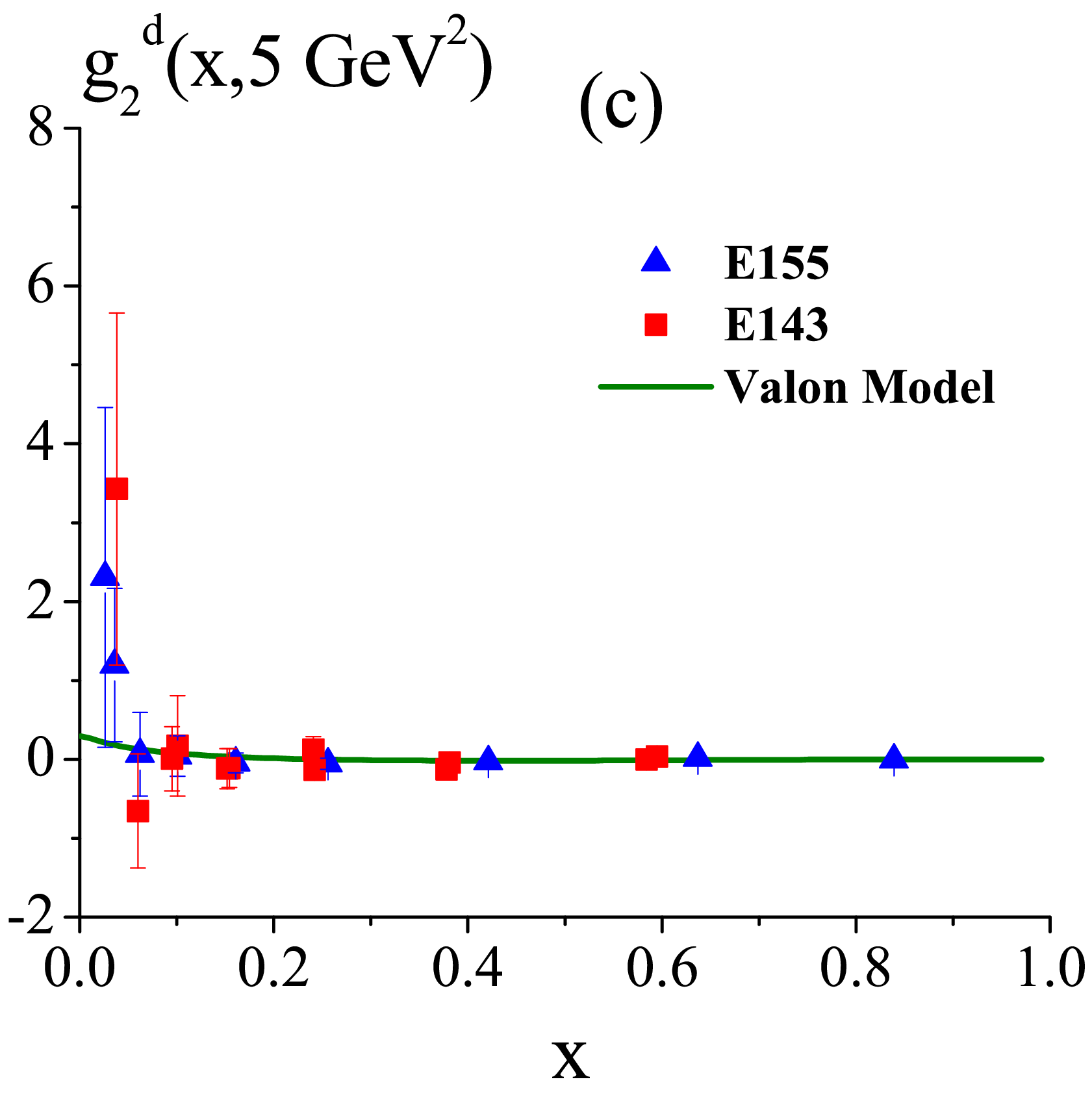}
\end{tabular}}
 \caption{\footnotesize  (Color online) Full transverse
polarized structure function $ g_2(x,Q^2)$ for proton, neutron and
deuteron at $Q^2 = 5\ GeV^2$ from valon model and it is compared
with experimental data \cite{abe,Anthony,HERMES}
 .}
\end{figure}

\subsection{The case of $g_{2}^{3_{He}}(x,Q^2)$}
 It is intriguing to investigate
the $g_{2}^{3_{He}}(x,Q^2)$ as a special case, since there are
some newly released data on $g_{2}^{3_{He}}(x,Q^2)$, and the
conformity of our result with the experiment will lend further
justification to the approach adopted here. \\

The $g_{2}^{3_{He}}(x,Q^2)$ structure function can be viewed as
the sum of $g_{2}^{n}(x,Q^2)$ and $g_{2}^{p}(x,Q^2)$, each
convoluted with the spin dependent nucleon light-cone momentum
distributions, $\Delta f_{3_{He}}^{N}(y)$, where $y$ is the ration
of "$+$ components of the light cone momenta of struck nucleon to
nucleus. One will have
\begin{equation}\label{he}
g_{2}^{3_{He}}(x,Q^2) = \int_{x}^{3}\frac{dy}{y}\Delta
f_{3_{He}}^{n}(y)g_{2}^{n}(x/y,Q^{2})+2
\int_{x}^{3}\frac{dy}{y}\Delta
f_{3_{He}}^{p}(y)g_{2}^{p}(x/y,Q^{2})
\end{equation}
So, using Eq.(\ref{he}), this should be straightforward. All is
needed are two functions, namely, $\Delta f_{3_{He}}^{p}(y)$ and
$\Delta f_{3_{He}}^{n}(y)$. They can be extracted from the
numerical results of \cite{Bissey,he3}. In Fig. 9 we have plotted
$g_{2}^{3_{He}}(x,Q^2)$ together
with the experimental data from \cite{E142,Jlab,Jlab2014}.
\begin{figure}[htp]
\centerline{\begin{tabular}{cc}
\includegraphics[width=7.5cm]{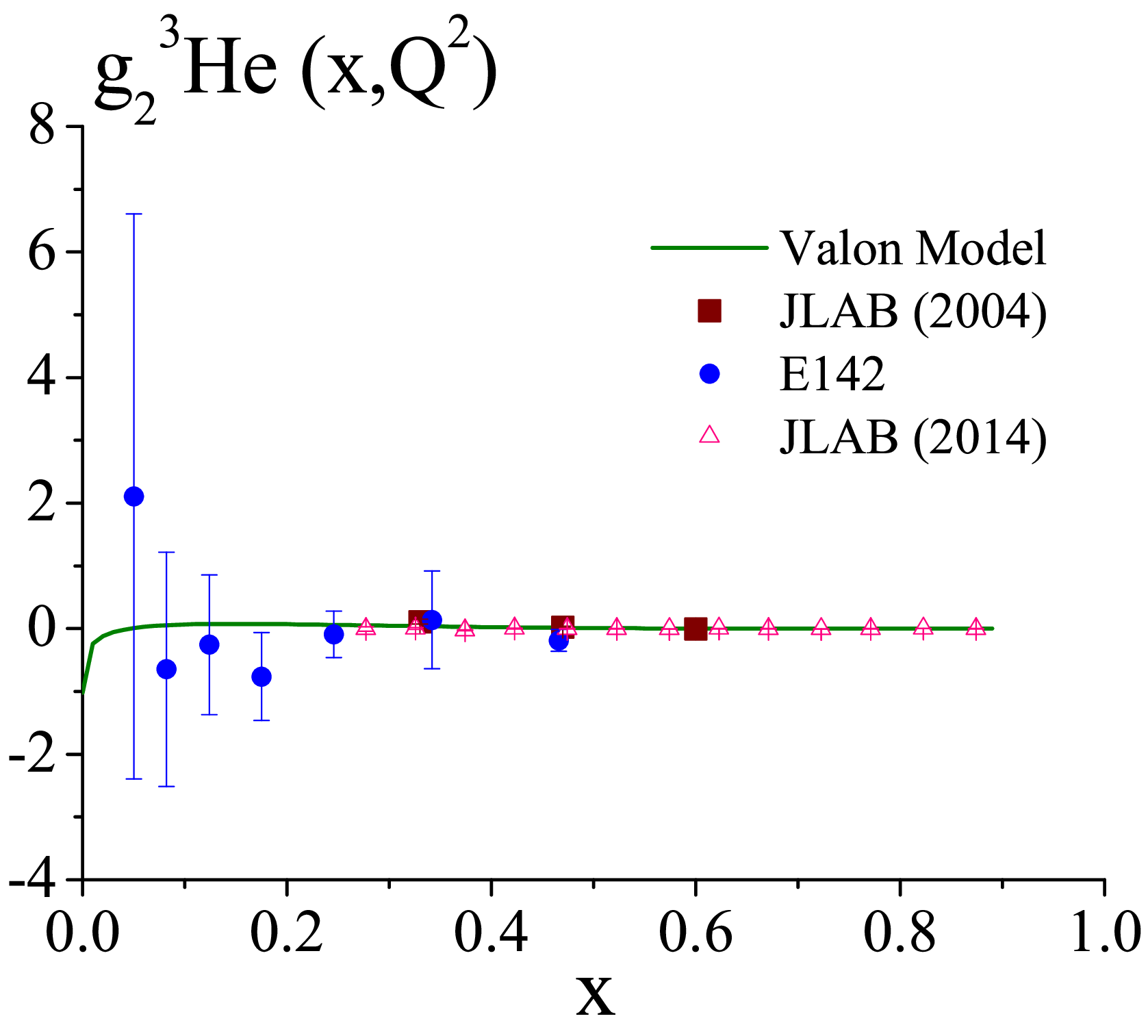}
\end{tabular}}
 \caption{\footnotesize (Color online) Full transversely polarized $3_{He}$
structure function, $g_2(x,Q^2)$  at $Q^2 =
5\ GeV^2$. The data points are from\cite{E142,Jlab,Jlab2014}.}
\end{figure}
As can be seen from the figure, our approach is in fair agreement with the experimental data.

\section{The Sum Rules}
 There are two important and well known sum rules regarding $g_{1}(x,Q^2)$ and  $g_{2}(x,Q^2)$. The first one is OPE sum rule:
\begin{equation}\label{gamma1}
\Gamma_1^{n}=\int_{0}^{1}x^n g_1(x,Q^2)dx=\frac{a_n}{2},
n=0,2,4,...
\end{equation}
\begin{equation}\label{gamma2}
\Gamma_2^{n}=\int_{0}^{1}x^n g_2(x,Q^2)dx=\frac{1}{2} \frac
{n}{n+1}(d_n-a_n), n=2,4,...
 \end{equation}
where $a_n$ and $d_n$ are the twist-2 and the twist-3 matrix
element operators, respectively. The study of these sum rules is
easy for simplest case (n=2) where the twist three effects are
exist \cite{shuryak1,shuryak2}.\\ The second one is the Burkhardt-
Cottingham sum rule \cite{Burkhardt}. It states that the first
moment of $g_{2}(x,Q^2)$ structure function vanished.
\begin{equation}
\int_{0}^{1} g_2(x,Q^2)dx=0
 \end{equation}
Since $g_{2}=g_{2}^{ww}+\bar g_{2}$, upon combining
Eq. (\ref{gamma1}) and Eq. (\ref{gamma2}) with Eq. (\ref{gww}) provides a third sum rule.
They are listed bellow:
\begin{eqnarray}
\int_{0}^{1} g_2^{ww}(x,Q^2)dx=0\\
\int_{0}^{1}x^2 g_2^{ww}(x,Q^2)dx=-\frac{1}{3}a_2\\
\int_{0}^{1}x^2 \bar g_{2}(x,Q^2)dx=\frac{1}{3}d_2
\end{eqnarray}
we have evaluated $d_{2}^{p},\ d_{2}^{n}$ in the valon model for a number
of $Q^{2}$ values, the results are shown in Fig. 10 and compared with the
available data, the bag model, and the QCD Sum Rule results. Table 3 shows
our results for the Burkhardt-Cottingham sum rule in the region
$0.023 < x < 0.9 $ at $Q^{2} = 5\ Gev^{2}$. They are checked against the data
from HERMES in the same region and also with the findings of E143 and E155 in
the range $0.02 < x < 0.8$. For the purpose of comparison,
results from other sources are also included. While $d_{2}^{p}$ is in
excellent agreement with the experiment, $d_{2}^{n}$ is less so. However,
we also notice that there are fewer data for $d_{2}^{n}$ and thus, making it
difficult to arrive in a firm conclusion.

\begin{table}
 {\footnotesize
\centerline{\begin{tabular}{|c|c|c|c|c|c|c|}
  \hline
   $$  & $a_2^p$ & $a_2^n$  & $a_2^d$   \\
  \hline
   \hline
Valon model &0.01956& -0.00004&0.00874 \\
  \hline
  Lattice QCD  \cite{Gockeler}& $(3\pm0.64)\times10^{-2}$ & $-(2.4\pm4.0)\times10^{-3}$ & $(13.8\pm5.2)\times10^{-3}$\\
   \hline
  CM bag model by Song  \cite{Song1}& $0.0210$  & $-1.8\times10^{-3}$& $0.0087$   \\
   \hline
  E143  \cite{abe}& $(2.42\pm0.20)\times10^{-2}$  & $-$& $(8.0\pm0.16)\times10^{-3}$  \\
   \hline
\end{tabular}}
 \caption{\label{a2} The twist-2 matrix elemet operators, $a_2$, for the proton, neutron and the deuteron, calculated in
the valon model. Also included the experimental data and the
results from other theoretical investigations. }}
\end{table}

\begin{table}
 {\footnotesize
\centerline{\begin{tabular}{|c|c|c|c|c|c|c|}
  \hline
   $$  & $d_2^p$ & $d_2^n$  & $d_2^d$   \\
  \hline
   \hline
Valon model & 0.00519 &0.0042&0.00437 \\
  \hline
  MIT bag model \cite{Song1,Ji} &$0.01$& $0$& $0.005$  \\
  \hline
  QCD sum rule  \cite{Stein}& $-(0.6\pm0.3)\times10^{-2}$  & $-(30\pm10)\times10^{-3}$ & $-0.017$\\
  \hline
  QCD sum rule  \cite{Balitsky}& $-(0.3\pm0.3)\times10^{-2}$  &$-(25\pm10)\times10^{-3}$& $-0.013$ \\
  \hline
  Lattice QCD  \cite{Gockeler}& $-(4.8\pm0.5)\times10^{-2}$ & $-(3.9\pm2.7)\times10^{-3}$ & $-0.022$ \\
   \hline
  CM bag model by Song  \cite{Song1}& $0.0174$  & $-2.53\times10^{-3}$& $0.0067$  \\
   \hline
  E143  \cite{abe}& $(0.54\pm0.5)\times10^{-2}$  & $-$& $(3.9\pm9.2)10^{-3}$ \\
   \hline
\end{tabular}}
 \caption{\label{d2} The twist-3 matrix elemet operators, $d_2$, for the proton, neutron and the deuteron,
calculated in the valon model. Also included the experimental data
and the results from other theoretical investigations. }}
\end{table}

\begin{figure}[htp]
\centerline{\begin{tabular}{cc}
\includegraphics[width=7.4cm]{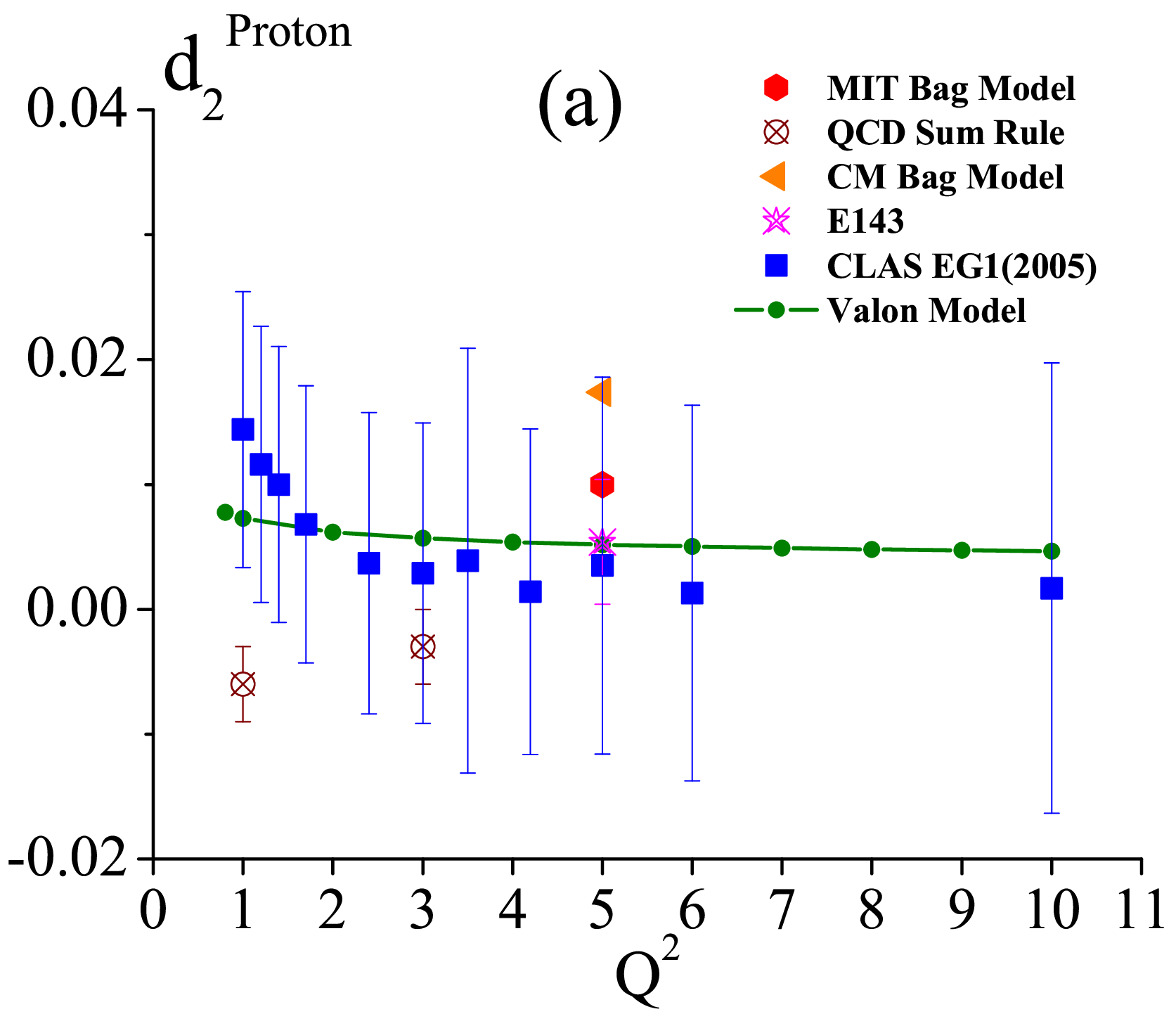}
\includegraphics[width=7.4cm]{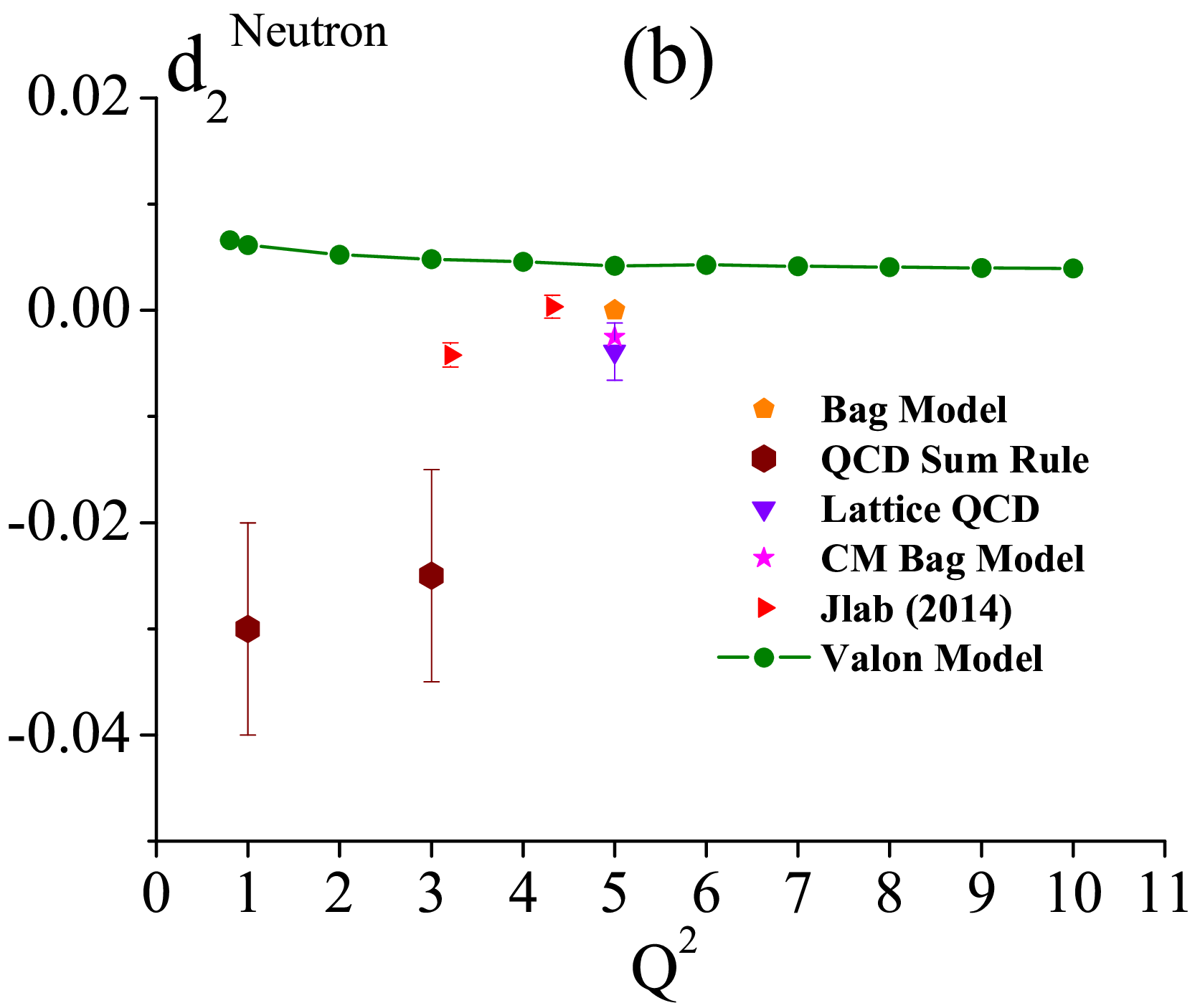}
\end{tabular}}
 \caption{\footnotesize  (Color online) $d_{2}^{p}\ and\ d_{2}^{n} $ are calculated in valon model.
Our results are compared with experimental data as well as with
obtained from  the results of other phenomenological groups,
 \cite{abe,Song1,Jlab2014,Ji,Stein,Balitsky,Gockeler,Osipko}
 .}
\end{figure}

\begin{table}
 {\footnotesize
\centerline{\begin{tabular}{|c|c|c|c|c|c|c|}
  \hline
   $$  & bag model by Song\cite{Song1} & E143 & E155 & HERMES 2012 & Valon model   \\
  \hline
   \hline
$\int_{0.02}^{0.85}g_2^p(x,Q^2)dx$ &-0.0016 & $-0.013\pm0.028$&$-0.022\pm0.071$&$0.006\pm0.024\pm0.017$ & -0.0016  \\
  \hline
  $\int_{0.02}^{0.85}g_2^d(x,Q^2)dx$ &-0.00287& ---&$0.023\pm0.044$&- & 0.0092 \\
  \hline
   \end{tabular}}
 \caption{\label{label} The results for the Burkhardt-Cottingham
sum rule. }}
\end{table}

\section{Conclusion}
We have used the so-called valon model to calculate the transverse
spin structure functions of the nucleon and the deuteron. To do
so, we provide a simple approach for calculating the twist-3 part
of the transverse spin structure function $\bar{g_{2}}(x,Q^{2)}$
in Mellin space. Furthermore, as a separate check on the validity
of our approach, we have considered $g_{2}^{3He}(x,Q^{2})$ , where
we have utilized some light cone momentum distribution and
compared with the new data from \cite{Jlab2014}. Evidently our
findings are in agreement with the experiment, rendering the
conclusion that hadronic structure functions, both polarized and
unpolarized, are nicely described in the valon representation.

\section*{Appendix }
Here we attempt to justify our choice of initial
input value in $g_2^{valon}(x,Q_0^2)$:
\begin{equation}
 \bar g_2^{valon}(z,Q_0^2) = A  \delta (z-1)\nonumber
\end{equation}

As we know the $g_2(x,Q^2)$ is related to quark-gluon-quark
correlation. Since, by definition, at initial scale, $Q_0^2$, the
valon behaves as an object with no internal structure,
it is reasonable to assume that, at such initial scale this object is related to the
quark-quark correlations(two-point green function), because at
such a low $Q_0^2$ gluons carry have very small and negligible momentum.\\
The general form of two-point green function in momentum space is
given here (Eq. (2.102) in \cite{book}):
\begin{eqnarray*}
 G^{(n)}(x_1,x_2,\cdots,x_n)&\cong&
 \frac{1}{2}(\frac{-i\lambda}{\hbar})\prod _{i=1}^{4} \hbar \frac{i}{p_i^2-m^2+i
 \epsilon} \int \frac{d^4k_1}{(2 \pi)^4}\frac{d^4k_2}{(2 \pi)^4}~(2
 \pi)^4 ~ \delta (p_1+p_2+k_1+k_2)\nonumber\\&&(2 \pi)^4 ~ \delta
 (p_3+p_4-k_1-k_2)~\hbar \frac{i}{k_1^2-m^2+i\epsilon} ~\hbar \frac{i}{k_2
 ^2-m^2+i\epsilon}
\end{eqnarray*}

The $g_2(x,Q^2 )$ structure function is related to the integral
over this two-point green function. We don't know the
exact relation between them, but at least we can say that it has two
terms: the first one is a function of $Q^2$ and the second one is
a Dirac Delta function which implies conservation of energy-momentum.
To establish this relation, we resort to the phenomenological arguments.
Obviously, the simplest choice
for $\bar g_2(x,Q^2)$ is
\begin{equation}
 \bar g_2^{valon}(z,Q^2) = f(Q^2) \delta (z-1) \nonumber
\end{equation}
at initial scale of $Q_0^2$, the photon probe detects only three
valence quarks inside the proton. Hence, we can assume that
$f(Q^2)\rightarrow A$ and we have:

\begin{equation}
 \bar g_2^{valon}(z,Q_0^2) = A  \delta (z-1) \nonumber
\end{equation}
The constant A can be determined from experimental data. Our
motivation for this value comes from the phenomenological
consideration which is required us to choose the initial input
densities as $\delta (z-1)$ at $Q_0^2$ . This mathematical
boundary condition means that the internal structure of the valon
cannot be resolved at $Q_0^2$. At this scale of $Q_0^2$ , the
nucleon can be considered as a bound state of three valence quarks
that carry all the momentum and the spin of the nucleon. As $Q^2$
is increased, other partons can be resolved at the nucleon.

\end{document}